\documentclass[12pt,a4paper]{article}
\usepackage{amsfonts}
\usepackage{amsmath}
\usepackage{amssymb}
\usepackage[english]{babel}
\usepackage{graphicx}
\newcommand{\tr}{\mathop{\rm Tr}\nolimits}
\begin{document}
\title{Quantum amplitudes are a consequence of elementary probability theory}
\author{Alexey L. Krugly\thanks{Quantum Information Laboratory, Institute of Physics and Physical Technologies, Moscow, akrugly@mail.ru}}
\date{} \maketitle
\begin{abstract}
I suppose that quantum objects obey elementary probability theory. I consider a connection of elementary probability theory and complex quantum amplitudes by a matrix calculus. A special case of a discrete pregeometry is an example of this approach.\end{abstract}
\section{Introduction}
Quantum mechanics is a mysterious scientific theory. It is a rigorous axiomatic theory and very successful in a research of processes in the microworld. An adequacy of quantum mechanics does not cause doubts. However a physical meaning of its base postulates remains not clear. Attempts to give an interpretation of quantum theory do not stop. A result of these efforts is a spectrum of interpretations (see e.g. \cite{SA} for an attempt to address this issue). This mystery of quantum mechanics is not solved.

A base quantity of quantum theory is a quantum amplitude. Therefore an interpretation of quantum mechanics means a physical interpretation of the quantum amplitude. The quantum amplitude is a complex number. A direct physical interpretation of a complex number is hardly possible. Only real numbers can have a direct physical meaning. Finally only non-negative integers can have a direct physical meaning of a scale reading of a device. I suppose that in quantum mechanics the complex numbers are only an effective calculation tool like a description of an alternating current in an electrical engineering. A base quantity is a probability of a discrete event. A physical meaning of this probability is a number of identical outcomes in a series of experiments. This is a non-negative integer.

Quantum mechanics is a probability theory of the microworld. But this theory is distinct from conventional probability theory. Probability theory is based on self-evident postulates. Therefore the existence of the alternative quantum probability theory is very mysteriously. Quantum mechanics can be formulated without complex amplitudes. We can use a sign-variable measure of a probability instead complex amplitudes \cite{RGV}. We can try to formulate laws for a non-negative quantum measure (see e.g. \cite{SS} for an attempt to address this issue). In this case there is a problem of an interpretation of negative probabilities or non-obvious rules for probabilities. In any case probabilities of events in the microworld do not submit to laws of probability theory. It is necessary to explain this fact.

Base laws of probability theory are formulated for statistically independent events. Probabilities can be connected by a non-obvious dependence if we consider statistically dependent events. I suppose that in quantum theory we have this situation. We are not able to divide a quantum process into statically independent events. For example, separate paths in a path integral are statistically dependent events. Quantum mechanics is very successful. Therefore quantum processes are adequately divided into structural components in quantum mechanics. But these structural components are statistically dependent. However I suppose that every quantum process can be divided into statistically independent components. Probability theory is fair for these components. Quantum mechanics is a non-obvious mathematical form of elementary probability theory.

It is possible to illustrate this idea by a following analogy. Let us consider an one-dimensional linear system. It consists of $N$ point masses. These masses are connected by ideal springs. Oscillations of any mass depend on oscillations of other masses and can be complicated. However we can divide any oscillations of the system into $N$ independent normal modes. The oscillations of each mass and the normal modes are two alternative descriptions. I suppose that a decomposition of quantum processes into statistically dependent structural components and statistically independent components is two alternative descriptions. We have the first variant. I suppose we can discover the second variant.

In the next section I describe this approach. A model of a continuum spacetime is not used. In section 3, I consider a particular case of a discrete pregeometry as an example. In section 4, there are a discussion of this approach. 

\section{Elementary probability theory and quantum amplitudes\label{QA}}
Consider some nonelementary event $X$. Assume that $X$ occur with a probability $p(X)$. Suppose $p(X)$ can be expanded into factors.
\begin{equation}
\label{eq:11} p(X)=(const)p_{ext}(X) p_{int}(X)\quad\textrm{,}
\end{equation}
where $const$ is a normalization constant, $p_{ext}$ depends on an external environment of $X$, and $p_{int}(X)$ depends on an internal structure of $X$. Consider only $p_{int}(X)$. Let
\begin{equation}
\label{eq:12} p_{int}(X)=2^{-I(X)}=\exp(-\ln(2)I(X))\quad\textrm{,}
\end{equation}
where $I(X)$ is a quantity of bits of an information that is contained in the structure $X$.

Assume that the internal structure of $X$ is a consequence of some discrete pregeometry. Suppose the internal structure of $X$ possesses $n$ statistically independent properties. Each property has a finite set of values. The number of the different values can be different for different properties. We have a set of $n$ fixed values for the structure $X$. One value for one property. This is a complete description of the structure $X$. Other properties do not influence on $p _ {int}(X)$. Therefore $p _ {int}(X)$ is a product of $n$ factors. Each factor is equal to a probability of the fixed value of one property. We have
\begin{equation}
\label{eq:13} p_{int}(X)=\exp(-\ln(2)\sum_{i=1}^n I(i, j))\quad\textrm{,}
\end{equation}
where $I(i_j)$ is a quantity of bits of an information that is contained in a property $i$ of the internal structure of $X$ if the property $i$ is equal to a value number $j$. By assumption, the number of statistically independent events is finite. Therefore the probability of $X$ is described by elementary probability theory.

In probability theory we can consider an arbitrary unnormalized probabilistic measure. Assume that a square matrix $\mathbf{X}$ of size $(n, n)$ describes the internal structure of $X$ and $I(i, j)$ is equal to the element $x_{ii}$ of  $\mathbf{X}$. Throughout matrixes will be designated by bold capital latin letters. This assumption can be a consequence of some matrix model. For example, such matrix models are discussed in \cite{SL08}. If the property $i$ is not equal to the value number $j$, $X$ has another structure and is described by another matrix of size $(n, n)$. We have
\begin{equation}
\label{eq:14} I(X)=\sum_{i=1}^n I(i, j)=\tr\mathbf{X}
\end{equation}
\begin{equation}
\label{eq:15} p_{int}(X)=\exp(-\ln(2)\tr\mathbf{X})\quad\textrm{.}
\end{equation}
Using a property of a matrix exponential, we get
\begin{equation}
\label{eq:16} p_{int}(X)=\exp(-\ln(2)\tr\mathbf{X})=\det\exp(-\ln(2)\mathbf{X})\quad\textrm{,}
\end{equation}
Consider a square complex matrix $\mathbf{A}$ of size $(n, n)$. By definition, put
\begin{equation}
\label{eq:17} \exp(-\ln(2)\mathbf{X})=\mathbf{A}\mathbf{A}^\dagger\quad\textrm{,}
\end{equation}
where $\mathbf{A}^\dagger$ is a hermitian transpose matrix. We can multiply $\mathbf{A}$ by an unitary matrix $\mathbf{U}$ of size $(n, n)$
\begin{equation}
\label{eq:18} \mathbf{A}\to\tilde{\mathbf{A}}=\mathbf{AU}\quad\textrm{.}
\end{equation}
$\tilde{\mathbf{A}}$ satisfies the equation (\ref{eq:17}). Assume that $\mathbf{A}$ is a hermitian matrix. We have
\begin{equation}
\label{eq:19} \mathbf{A}^2=\exp(-\ln(2)\mathbf{X})
\end{equation}
\begin{equation}
\label{eq:110} \mathbf{A}=\exp(-2^{-1}\ln(2)\mathbf{X})\quad\textrm{.}
\end{equation}
We can get any nonsingular matrix $\tilde{\mathbf{A}}$ that satisfies (\ref{eq:17}) by the transformation (\ref{eq:18}) if $\mathbf{A}$ satisfies the equation (\ref{eq:110}). Using a similarity transformation, we have
\begin{equation}
\label{eq:111}
\begin{array}{rcl}
p_{int}(X)=\det\exp(-\ln(2)\mathbf{X})&=&\det(\mathbf{U}^\dagger\exp(-\ln(2)\mathbf{X})\mathbf{U})=\\
&&=\det\exp(-\ln(2)\mathbf{U}^\dagger\mathbf{X}\mathbf{U})\quad\textrm{.}
\end{array}
\end{equation}
The equality in the right side of (\ref{eq:111}) is a property of matrix functions. Using (\ref{eq:17}) and (\ref{eq:18}) we have
\begin{equation}
\label{eq:112} p_{int}(X)=\det\exp(-\ln(2)\mathbf{U}^\dagger\mathbf{XU})=\det\mathbf{U}^\dagger\mathbf{AU}\det\mathbf{U}^\dagger\mathbf{A}^\dagger\mathbf{U}\quad\textrm{.}
\end{equation}
$\det\mathbf{A}$ is called an amplitude of the event $X$. $\det\mathbf{A}\det\mathbf{A}^+$ is invariant under the transformations (\ref{eq:18}) and (\ref{eq:112}). A matrix function is a sum of infinite matrix series. Using (\ref{eq:110}), we get
\begin{equation}
\label{eq:113} \mathbf{A}=\exp(-2^{-1}\ln(2)\mathbf{X})=\sum_{n=0}^{\infty}\frac{(-\ln(2)\mathbf{X})^n}{2^nn!}\quad\textrm{.}
\end{equation}
$\det\mathbf{A}$ is the sum of the infinite number of alternating summands. We can get $\det\mathbf{A}$ as the sum of the infinite number of complex summands by the transformations (\ref{eq:18}) or (\ref{eq:112}). We get the sum of complex amplitudes of alternatives if we regard these summands as alternatives. Such calculation of probabilities seems mysterious only if we consider these \flqq alternatives\frqq\ as statistically independent alternatives.

Quantum systems are classical stochastic systems in the considered model. They have no a deterministic dynamics. A different approach is offered in \cite{SL02} where quantum laws are a consequence of a classical statistical description of a matrix model. However a dynamics of a system is deterministic and a stochastic description is secondary, similarly a classical statistical theory.

Finally, consider a property of independent events. Let $\mathbf{X}_{1 2}$ be a matrix of two independent events 1 and 2. Assume that
\begin{equation}
\label{eq:114} \mathbf{X}_{1 2}=\mathbf{X}_1+\mathbf{X}_2\quad\textrm{.}
\end{equation}
We have
\begin{equation}
\label{eq:115} \tr\mathbf{X}_{1 2}=\tr\mathbf{X}_1+\tr\mathbf{X}_2\quad\textrm{,}
\end{equation}
Using (\ref{eq:14}), we get
\begin{equation}
\label{eq:116} I(X_{1 2})=I(X_1)+I(X_2)\qquad\textrm{.}
\end{equation}
An amount of an information of two independent events is a sum of amounts of an information of first and second events. We have
\begin{equation}
\label{eq:117}
\begin{array}{rcl}
\mathbf{A}_{1 2}=\exp(-2^{-1}\ln(2)\mathbf{X}_{1 2})=\exp(-2^{-1}\ln(2)(\mathbf{X}_1+\mathbf{X}_2))&=&\\
=\exp(-2^{-1}\ln(2)\mathbf{X}_1)\exp(-2^{-1}\ln(2)\mathbf{X}_2)&=&\mathbf{A}_1\mathbf{A}_2
\end{array}
\end{equation}
\begin{equation}
\label{eq:118} \det\mathbf{A}_{1 2}=\det\mathbf{A}_1\det\mathbf{A}_2\quad\textrm{.}
\end{equation}
An amplitude of two independent events is a product of amplitudes of first and second events.

\section{A model of a pregeometry\label{MP}}
Consider a special case of a discrete pregeometry as an example of this approach. A model of the pregeometry is a finite directed acyclic graph. A graph is a set of vertexes and a binary relation over this set. The vertexes are denoted by lowercase latin letters $a$, $b$, $\dots$ The binary relation $(a b)$ over the set of vertexes is called an edge. All edges of a directed graph are directed (we take into account the order of the vertexes in $(a b)$). A subset of vertexes is called a sequence if every two neighboring vertexes are connected by an edge. A sequence is a cyclic sequence if an initial vertex and a final vertex coincide. A sequence is called a directed sequence if any two consecutive vertexes are an origin and an end of a common edge. A sequence is called an opposite directed sequence if any two consecutive vertexes are an end and an origin of a common edge. A directed graph is called a directed acyclic graph if it contains no directed cyclic sequences.

In this model the vertexes of the graph are elementary events like spacetime points. The edge $(a b)$ has physical meaning of an elementary cause relation between the events $a$ and $b$. A physical meaning of a directed sequence is a cause-effect relation. The initial vertex $a$ of a directed sequence is a cause and the final vertex $b$ of this sequence is an effect.

The cause-effect relation of vertexes is an order-relation. We represent the order-relation by $\preceq$ and use the reflexive connection that a vertex precedes itself. A set of vertexes of a finite directed acyclic graph is a locally finite partially ordered set (or \flqq poset\frqq) or a causal set (or \flqq causet\frqq) \cite{Myrheim, BMSorkin, Sorkin1}. However a base relation of this graph is the directed edges. This is a relation of an immediate causal priority \cite{Fink88}.

Let us assume following terms for a causal set $C$ of vertexes. The past of an vertex $a$ is the subset $past(a)=\{ b\in C|(b\preceq a)\}$. This is the past light cone of $a$. The vertex of $C$ is maximal if it is to the past of no other vertex. The future of an vertex $a$ is the subset $future(a)=\{b \in C|(a\preceq b)\}$. This is the future light cone of $a$. The vertex of $C$ is minimal if it is to the future of no other vertex. A directed sequence is an inextendible directed sequence if every vertex not in it is not related by a cause-effect relation to some vertex of this sequence. An initial vertex of an inextendible directed sequence is a minimal vertex. A final vertex of an inextendible directed sequence is a maximal vertex.

Every finite graph can be defined by an incidence matrix or a vertex incidence matrix or an edge incidence matrix. A vertex incidence matrix is called an adjacency matrix. We can describe properties of a graph as operations on matrixes. Therefore we can construct an example of the considered approach for this pregeometry.

Consider an adjacency matrix $\mathbf{V}$. The element $v_{ij}$ of $V$ is equal to zero if there is no the edge $(i j)$. The element $v_{ij}$ is equal to one if there is one edge $(i j)$. The element $v_{ij}$ is equal to the number of the edges $(i j)$ if there are the multiple edges $(i j)$. $\mathbf{V}$ is a square matrix. The size of $\mathbf{V}$ is equal to the number of vertexes in the graph.

Let $n$ be the number of maximal vertexes and $m$ be the number of minimal vertexes. Consider a matrix $\mathbf{S}$. The element $s_{ij}$ of $\mathbf{S}$ is equal to the number of inextendible directed sequences from the minimal vertex $i$ to the maximal vertex $j$. $\mathbf{S}$ is a rectangular matrix of size $(m, n)$. We can get $\mathbf{S}$ using $\mathbf{V}$. The element $v_{ij}(k)$ of $\mathbf{V}^k$ is equal to the number of directed sequences from the vertex $i$ to the vertex $j$ such that every sequence contains $k$ edges. Therefore the number of directed sequences from the vertex $i$ to the vertex $j$ is equal to the element $w_{ij}$ of a matrix $\mathbf{W}$:
\begin{equation}
\label{eq:21} \mathbf{W}=\sum_{k=0}^{N}\mathbf{V}^k\quad\textrm{,}
\end{equation}
where $N$ is equal to the number of edges in the graph. The last summand in the sum (\ref{eq:21}) is not a zero matrix if the graph is a directed sequence. We get $\mathbf{S}$ from $\mathbf{W}$ by deletion its rows and columns if the row is not corresponded to a minimal vertex and if the column is not corresponded to a maximal vertex.

\begin{figure}[ht]
	\centering	
		\includegraphics[width=5cm,trim=8cm 15cm 8cm 3cm]{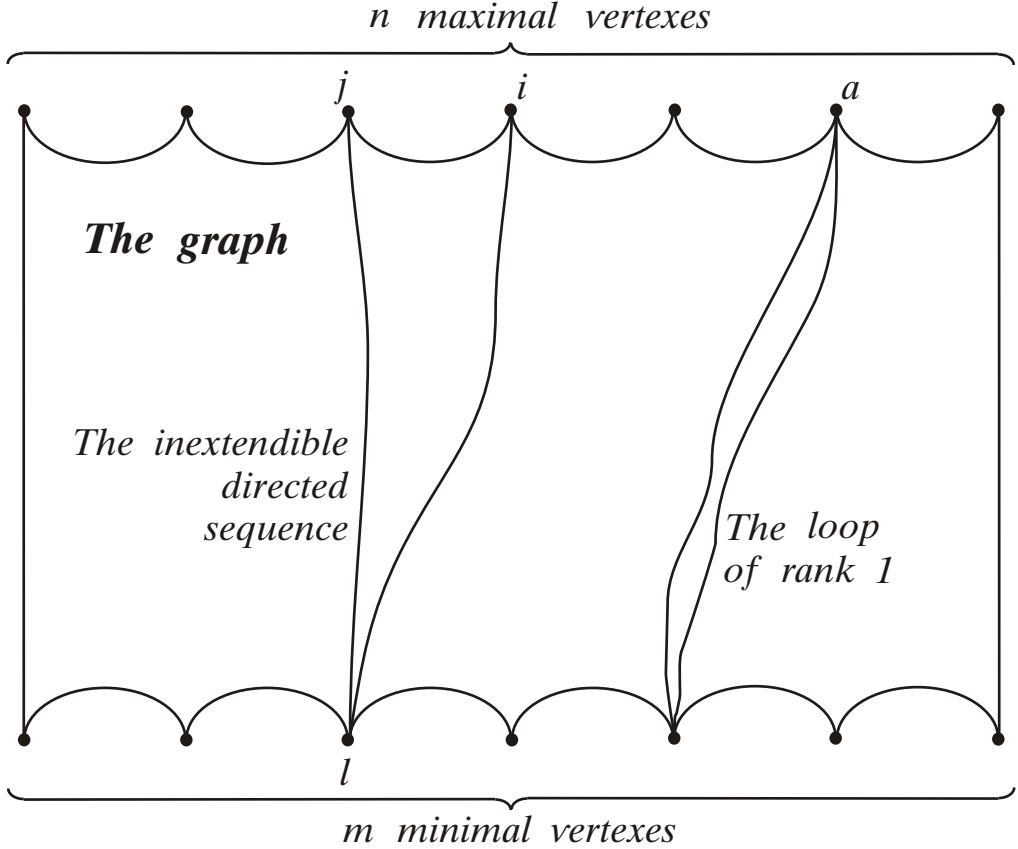}
	\caption{These sequences are corresponded to the elements of $\mathbf{S}^T\mathbf{S}$.}
	\label{fig:fig1}
\end{figure}

The element $s^T_{ij}$ of the transposed matrix $\mathbf{S}^T$ is equal to the number of opposite directed sequences from the maximal vertex $i$ to the minimal vertex $j$. The element $s_{ij}(2)$ of $\mathbf{S}^T\mathbf{S}$ is equal to the number of sequences that consist of two parts (fig.~\ref{fig:fig1}). The first part is an opposite directed sequences from the maximal vertex $i$ to some minimal vertex $l$. The second part is a directed sequences from the minimal vertex $l$ to the maximal vertex $j$. We have a summation over vertexes $l$ in the element $s_{ij}(2)$. The diagonal element $s_{aa}(2)$ is equal to the number of all cyclic sequences from the maximal vertex $a$ to each minimal vertex and back (fig.~\ref{fig:fig1}). First part of such sequence is an opposite directed sequence from $a$ to a minimal vertex. Second part of such sequence is a directed sequence from this minimal vertex to $a$. Let us call such cyclic sequence a loop of rank 1. The size of $\mathbf{S}^T\mathbf{S}$ is $(n, n)$.
Assume that $\mathbf{S}^T\mathbf{S}$ is $\mathbf{X}$. Using (\ref{eq:115}), we have
\begin{equation}
\label{eq:22} p_{int}(X)=\exp(-\ln(2)\tr\mathbf{S}^T\mathbf{S})\quad\textrm{.}
\end{equation}
Consider a physical meaning of (\ref{eq:22}). $\tr\mathbf{S}^T\mathbf{S}$ is equal to the number of loops of rank 1 in the graph. Therefore each loop of rank 1 contains one bit of information about the internal structure of the graph. In other words each loop of rank 1 is a realization of one outcome of some binary alternative. Binary alternatives are discussed in the section 44.5 of \cite{gr} as a background of a dynamics of the microworld (also see e.g. \cite{WFG, FinkMcC1, Weiz1, FinkMcC2, Weiz2,
Weiz3}). This list of references is by no means complete. The identification of a loop of rank 1 with a binary alternative should be a consequence of some fundamental principle. This principle can constrain the structure of the graph.

Consider a statistical assembly that is a set of graphs. Each graph contains $m$ minimal vertexes, $n$ maximal vertexes and $M$ loops of rank 1. Each graph is a simple event of a sample space. This simple event is a joint realization of $M$ binary alternatives. $p_{int}(X)$ is proportionate to $2^{-M}$. We can get the normalization constant by a summation over probabilities of simple events. A treatment of another statistical assemblies is possible. A choice of a statistical assembly depends on a considered problem, initial and boundary conditions. This choice is related to an identification of a properties of a graph and observables.

$M$ multipliers $2^{-1}$ are integrated in (\ref{eq:22}) to $n$ groups of multipliers. The number of groups is the number of maximal vertexes. The number of multipliers in each group is the number of loops of rank 1 in a past light cone of a maximal vertex. In other words $s_{aa}(2)$ equal to the number of bits of information that is contained in the past light cone of the maximal vertex $a$. $s_{aa}(2)$ is determined only by the structure of $past(a)=\{ b\in C|(b\preceq a)\}$. This is a form of a causality principle.

We can permute $\mathbf{S}^T$ and $\mathbf{S}$ in (\ref{eq:22}). The size of $\mathbf{SS}^T$ is $(m, m)$. In this matrix, $M$ multipliers $2^{-1}$ are integrated to $m$ groups of multipliers. The number of groups is the number of minimal vertexes. In this case, a dynamics is determined by structures of future light cones of minimal vertexes. Therefore a time reversal is possible in the considered model. This is a reversal of directions of all edges in the graph.

\begin{figure}[ht]
	\centering	
		\includegraphics[width=5cm,trim=8cm 15cm 8cm 3cm]{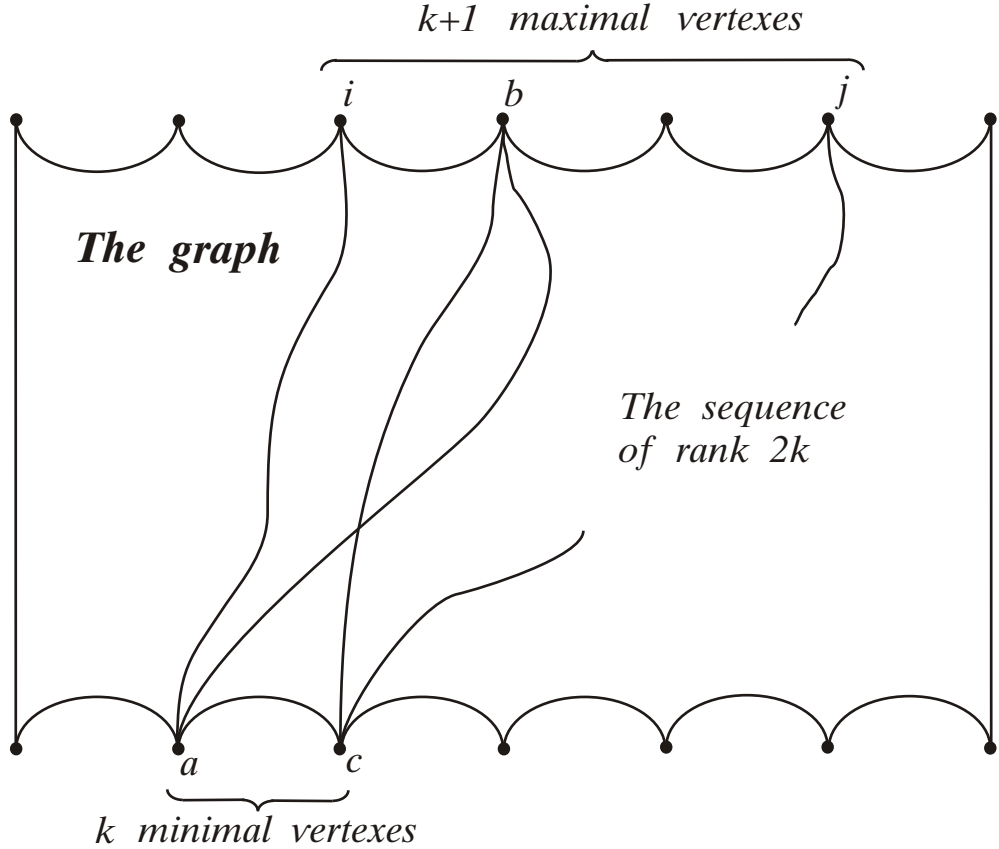}
	\caption{The sequence of rank $2k$.}
	\label{fig:fig2}
\end{figure}

Consider a quantum amplitude of the graph. Using (\ref{eq:110}) and (\ref{eq:22}), we get
\begin{equation}
\label{eq:23} \det\mathbf{A}=\det\exp(-2^{-1}\ln(2)\mathbf{S}^T\mathbf{S})\quad\textrm{.}
\end{equation}
Using (\ref{eq:113}), we have
\begin{equation}
\label{eq:24} \mathbf{A}=\sum_{k=0}^{\infty}\frac{(-\ln(2)\mathbf{S}^T\mathbf{S})^k}{2^kk!}\quad\textrm{.}
\end{equation}
The element $s_{ij}(2, k)$ of $(\mathbf{S}^T\mathbf{S})^k$ is the number of following sequences (fig.~\ref{fig:fig2}). Each sequence consist of $k$ inextendible opposite directed sequences and $k$ inextendible directed sequences. It starts in the maximal vertex $i$, includes an inextendible opposite directed sequence from $i$ to some minimal vertex $a$, then includes an inextendible directed sequences from $a$ to some maximal vertex $b$, then includes an inextendible opposite directed sequence from $b$ to some minimal vertex $c$ and so on. The last part is an inextendible directed sequences to the maximal vertex $j$. Such sequence is called a sequence of rank $2k$. If $i$ and $j$ coincide a sequence of rank $2k$ is called a loops of rank $k$. The element $a_{ij}$ of $\mathbf{A}$ is an infinite sum of the numbers of sequences of all even ranks from the maximal vertex $i$ to the maximal vertex $j$ such that each sequence is multiplied by some factors.

The size of $\mathbf{A}$ is $(n, n)$, where $n$ is the number of maximal vertexes. $\det\mathbf{A}$ is the sum of $n!$ summands. Each summand is a product of $n$ different elements of $\mathbf{A}$, where there is one element of each row and each column. This is a product of $n$ infinite series. Consequently $\det\mathbf{A}$ is an infinite series. An end of any sequence is an origin of other sequence in each summand of this infinite series. Therefore $\det\mathbf{A}$ is an infinite sum over loops. Different loops can have different coefficients and different inextendible directed sequences in the same loop can have different coefficients.

We can get a complex quantum amplitude of the graph by the transformations (\ref{eq:18}) or (\ref{eq:112}). In this case, different inextendible directed sequences in the same loop can have different complex coefficients.

A calculation of $\det\mathbf{A}$ as this sum of the infinite series is a very difficult problem for big graphs. However in any case, we have
\begin{equation}
\label{eq:25} (\det\mathbf{A})(\det\mathbf{A})^*=2^{-M}\quad\textrm{,}
\end{equation}
where $M$ is the number of binary alternatives in the considered graph.

Two independent events are described by a disconnected graph that consists of two connected subgraphs. In this case, we can choose a numbering of vertexes such that $\mathbf{S}^T\mathbf{S}$ is a block-diagonal matrix. Each block corresponds to the subgraph. A trace of a block-diagonal matrix is equal to the sum of traces of blocks. This is consistent with (\ref{eq:114}) - (\ref{eq:118}).

Two sequential phases of an evolution of a system are described by two subgraphs $X_1$ and $X_2$. Maximal vertexes of $X_1$ coincide with minimal vertexes of $X_2$. Each inextendible directed sequence of the graph consist of an inextendible directed sequences of $X_1$ and an inextendible directed sequences of $X_2$. Therefore
\begin{equation}
\label{eq:26} \mathbf{S}=\mathbf{S}_1\mathbf{S}_2\quad\textrm{,}
\end{equation}
$\tr(\mathbf{S}^T\mathbf{S})$ is not equal to $\tr(\mathbf{S}^T_1\mathbf{S}_1)+\tr(\mathbf{S}^T_2\mathbf{S}_2)$. The model describes a statistically dependence of sequential phases of an evolution of a system.
\section{Discussion}
The considered example shows that a calculus of quantum amplitudes can be a mathematical form of elementary probability theory.

In this model an evolution of a system is described by (\ref{eq:26}). This is \flqq an algebra of directed sequences\frqq. Probabilities depend on loops. Each inextendible directed sequence is included in loops twice: first time in an opposite direction and second time in a straight direction. This determine the square-law of an information on $\mathbf{S}$. Loops are \flqq a square\frqq\ of directed sequences. Consequently a probability theory of a pregeometry is \flqq a square of a dynamics\frqq.

We can formally identify $\mathbf{X}$ with the square matrix $\mathbf{S}$ of a graph with equal numbers of maximal and minimal vertexes. However this assumption has not a physical meaning. $\tr\mathbf{S}$ is equal to the number of inextendible directed sequences from minimal vertexes to maximal vertexes that have equal numbers. $\tr\mathbf{S}$ depends on a numbering of vertexes. This is unacceptable.

In this model the quantum amplitude is always a determinant. In practical problem the amplitude can be different from a determinant. First we can sum over finite part of the infinite series (\ref{eq:113}). Secondly any summands can be a negligible quantity. Thirdly some summands can join together. This reflects association of edges and vertexes in elementary particles.

The transformations (\ref{eq:18}) and (\ref{eq:112}) must have the physical meaning for the adequate model. An interesting case is a Jordan canonical form of $\mathbf{X}$.

We can consider different generalizations of the model. $\mathbf{S}^T\mathbf{S}$ includes confluent loops. This loop consists of one inextendible directed sequence that is included twice, one time in an opposite direction and second time in a straight direction. We can take away them by diminution of $\mathbf{S}^T\mathbf{Z}$ from $\mathbf{S}^T\mathbf{S}$. $\mathbf{Z}$ is a matrix of size $(m, n)$. All elements of $\mathbf{Z}$ is equal to 1. $\mathbf{S}^T\mathbf{S}$ includes each loop twice, one time in one direction and second time in an opposite direction. We can take away this effect by a division of $\mathbf{S}^T\mathbf{S}$ by 2. We can consider another kinds of loops. For example, these are loops without a repeating of edges or vertexes. In this case, we can use suitable mathematical tools \cite{K1, K2, K3, K5}. We assign a generator of a one-dimensional Grassmann algebra to each edge (or vertex). Elements of the adjacency matrix are these generators or their products. We have powers of the generators in products of the adjacency matrixes. They are equal to zero. The power of the generator is equal to the number of a repeating of an edge or a vertex in a sequence. We get an exception of such sequence. Integrating the matrix with respect to all generators, we obtain a matrix such that elements are equal to the numbers of sequences without repeating of edges or vertexes. We can consider a model with weighted edges or vertexes. Therefore a generalizations of the model can describe different kinds of a dynamics.

The equations (\ref{eq:23}) and (\ref{eq:24}) must correspond to a sum over Feynman diagrams for an adequate model. In this case, edges and vertexes must correspond to some matrixes. We can consider a coarse graining of the graph. Meta-edges and meta-vertexes of an aggregated graph can correspond to subgraphs of the initial graph and can be described by matrixes. Another possibility is an extension of the transformations (\ref{eq:18}) and (\ref{eq:112}). Elements of the matrix $\mathbf{U}$ can be matrixes. Feynman diagrams have external lines. A graph can not have external edges because an edge is a binary relation of two vertexes by a definition. However we can consider a different mathematical structure. Consider edges as primary units and vertexes as relations of edges \cite{K2, K3, K4, K5}. External edges can exist in this structure. Consider a set of directed edges and define the order-relation. External edges are minimal or maximal elements. We can consider an incidence matrix and we can develop this approach for such structure.

A pregeometry must describe all: spacetime and a matter. In causal set theory discrete elements are uniformly distributed in spacetime \cite{BMSorkin}. In this case, we can describe empty spacetime but we must add matter ad hoc. A pregeometry must form nonuniform hierarchical structures. Loops of a directed graph can be such structures.

There are cause-effect connections only. Any simultaneous structure is a set of disconnected points. These points are connected by intersections of past light cones. Consequently any topological structure is a process. The loops are topological objects. They are not local objects of the graph. A topological model of particles is offered in \cite{BTSO1}. This is two sets of vertexes and braided connections between these sets. This object can be a simultaneous structure \cite{BTSOMFSL1, 0804.0037}. However in the considered model any structures are processes. Such topological objects can be one cycle of a periodical process. One set of vertexes is an initial state, second set of vertexes is a final state and braided connections are the cycle of the process. The same vertexes are the final state of the previous cycle and the initial state of the next cycle.

In statistical physics there is a state probability at a time. A system evolves to most probable state. In the considered model there is a probability of an evolutionary process. A system evolves with high probability of the evolutionary process. {\itshape In this case, the evolutionary process has a minimum quantity of an information. This can be a form of a principle of least action and a quant of an action is a bit of an information.}

In the considered model the binary alternatives are the loops in a graph. A structure of binary alternatives must be a consequence of fundamental principles in an adequate model of the microworld. This fundamental principles are a subject of further investigation. This and related issues are sure to determine the directions of future research.

I am very grateful to my wife Valeria G. Koshelayeva for a technical assistance.

\end{document}